\documentclass[10pt,a4paper]{article}
\usepackage{amssymb}
\usepackage{amsmath}
\usepackage{epsfig}

\newcommand{\ket}[1]{| #1 \rangle}
\newcommand{\bra}[1]{\langle #1 |}

\begin{document}
\title{Comment on ``Experimental realization of Popper's experiment: Violation of the uncertainty
principle?''}
\author{A.J.Short\thanks{tony.short@qubit.org}\\ {\em Centre for Quantum Computation,} \\
{\em Clarendon Laboratory, University of Oxford,}\\ {\em Parks
Rd., OX1 3PU, UK}}
\date{}
\maketitle{}

\begin{abstract}
Application of the uncertainty principle to conditional
measurements is investigated, and found to be valid for
measurements on separated sub-systems. In light of this, an
apparent violation of the uncertainty principle obtained by Kim
and Shih in their realization of Popper's experiment
\cite{main_paper} is explained through analogy with a simple
optical system.
\end{abstract}

\section{Introduction}

The entangled photon pairs produced in spontaneous parametric
down-conversion (SPDC) have been used in a range of experiments to
demonstrate non-local correlations in quantum mechanics. In this
paper we examine the results of one such experiment, in which
Yoon-Ho Kim and Yanhua Shih obtain an apparent violation of the
uncertainty principle \cite{main_paper}.

Their experiment is the modern realization of a thought experiment
created by Karl Popper in the early 1930's to illustrate his
doubts about the uncertainty principle
\cite{main_paper}\cite{popper}\cite{asher_view}. Although his
proposal was flawed because of its reliance on a stationary point
source, which is itself forbidden by the uncertainty principle
\cite{popper_flaw}, Kim and Shih have recreated the essence of
Popper's experiment without this problem using SPDC and a
converging lens. Surprisingly, their experimental data appears to
show a violation of the uncertainty principle in agreement with
Popper's original prediction.

After a review of Kim and Shih's experiment, we investigate the
validity of the uncertainty principle when dealing with entangled
systems and conditional measurements. In light of this, we then
look more closely at their experiment and offer an alternative
explanation for the results. As with Popper's original experiment,
we will show that the width of the source plays a crucial role.

\section{The experiment}

\begin{figure}[b]
\centerline{ \includegraphics[width=12cm]{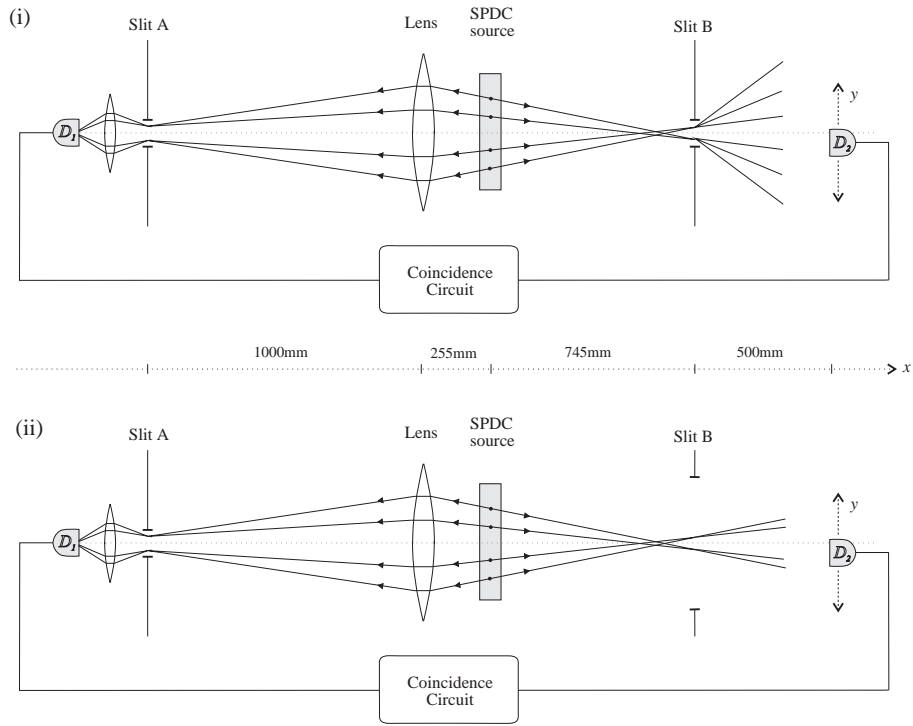} }
\caption{\small The ``unfolded'' schematic of Kim and Shih's
experiment, showing the two cases investigated. In case (i) slit B
is narrowed to the same width as slit A (0.16mm), while in case
(ii) slit B is wide open. Experimental results show that the
momentum spread $\Delta p_y$ of the photon passing through slit B
is almost three times as large when the slit is narrowed.}
\label{unfolded_diagram}
\end{figure}

As explained in \cite{main_paper}, Kim and Shih's experiment is
conceptually equivalent to the ``unfolded'' schematic given in
Fig. \ref{unfolded_diagram}, in which an SPDC source, lens and
slits all lie on a common $x$-axis. The central converging lens
(of focal length 500mm) is positioned 255mm from the SPDC source
and 1000mm from each of two parallel slits (A and B), with slit B
closest to the source. Pairs of entangled photons are generated at
the source by an external pump-laser, and their trajectories are
measured on either side of the slits by detectors $D_1$ and $D_2$.
A collecting lens is used to channel light passing through slit A
into the fixed detector $D_1$, while $D_2$ is scanned along the
$y$-axis 500mm behind slit B. The results are then combined in a
coincidence circuit to yield a conditional measurement:\emph{The y
coordinate of photon 2 in the plane of $D_2$ given that photon 1
passes through slit A and is detected at $D_1$.} Two cases are
studied; (i) in which slit B is the same width as slit A (0.16mm),
and (ii), in which slit B is wide open.

Although the photons are created with a large uncertainty in
position and momentum there are strong correlations between each
pair due to the phase matching conditions of SPDC. In this
``unfolded'' schematic the momenta of the photons in each pair
($\hbar \mathbf{k}_1$ and $\hbar \mathbf{k}_2$)\footnote{Photons 1
and 2 are often referred to as `signal' and `idler' photons
respectively.} are almost precisely anti-correlated, with
$\mathbf{k}_1+\mathbf{k}_2 \simeq 0$. Because of this, the
two-photon trajectories are well represented by straight lines and
may be treated like optical rays\footnote{This property is the
major motivation for ``unfolding'' the experiment in this way, the
procedure for which is explained more fully in refs.
\cite{main_paper} and \cite{image_expt}}.

\begin{figure}[b]
\centerline{ \includegraphics[width=12cm]{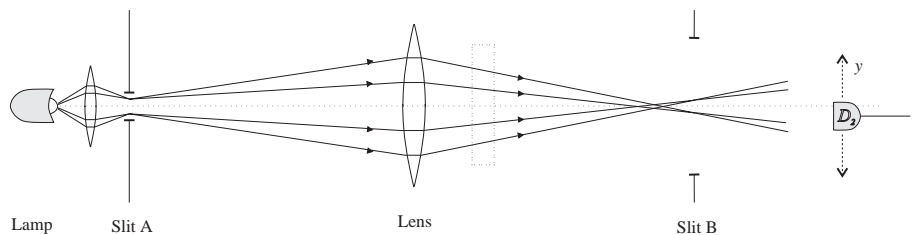} }
\caption{\small The schematic for an analogous single-photon
experiment, in which $D_1$ is replaced by a lamp and the SPDC
source removed. Note the similarity to figure
\ref{unfolded_diagram} and the change in the direction of
propagation of photons to the left of the lens. }
\label{simple_optical_fig}
\end{figure}

We can construct a single-photon system which will generate
approximately the same results as this experiment by replacing
detector $D_1$ with a lamp and removing the SPDC source, as shown
in Fig. \ref{simple_optical_fig}. Individual photons then
propagate from the source at $D_1$ through the collection lens,
slit A, the central lens, and slit B before being detected at
$D_2$. As discussed in \cite{single_photon1} and
\cite{single_photon2}, results for this simple optical setup will
be very similar to coincidence measurements in the SPDC
experiment. In particular, we expect a ``ghost image'' of slit A
to be observed in coincidence measurements on the photon pair,
just as a conventional image is visible in the single-photon
setup, and this has been verified experimentally by Pittman
\emph{et al.} \cite{image_expt}.

By positioning both slits two focal lengths (1000mm) away from the
lens, we can produce an unmagnified image of slit A in the plane
of slit B\footnote{This follows from the Gaussian thin lens
equation $\frac{1}{a} + \frac{1}{b} = \frac{1}{f}$, where $a$ and
$b$ are the positions of slit A and its image relative to the
lens, and $f$ is the focal length. $a=b=2f$ gives an unmagnified
solution.}. If photon 1 passes through slit A and is detected at
$D_1$ then photon 2 must pass through the image in the plane of
slit B, as shown in fig. \ref{unfolded_diagram}. The image slit
should be the same width as slit A (0.16mm), so one would not
expect the behaviour of photon 2 to be affected if slit B is also
narrowed to this width. However, according to Kim and Shih's
results \emph{the momentum spread of photon 2 when slit B is
narrowed is almost three times that when slit B is wide open.} It
appears that the presence of a physical slit affects the results
even though it does not change the spatial confinement of the
photon.

The momentum uncertainty of photon 2 in the image plane can be
deduced approximately from the width of its spatial distribution
at $D_2$. In case (i), with slit B narrowed to the same width as
slit A (0.16mm), the distribution at $D_2$ is $4.4$mm wide.
However, when slit B is wide open in case (ii) the width at $D_2$
is reduced to 1.6mm. Using simple geometrical arguments we obtain
$\Delta_{(ii)} p_y \approx 0.36 \Delta_{(i)} p_y$, where
$\Delta_{(i,ii)}$ refers to the uncertainty in cases (i) and (ii)
respectively. If we accept the above arguments and take the
uncertainty in $y$ to be the same in both cases ($\Delta_{(i)} y =
\Delta_{(ii)} y = 0.16$mm) then there is significant reduction in
$\Delta y \Delta p_y$, which suggests a violation of the
uncertainty principle\footnote{It is difficult to obtain a
definite violation of the uncertainty principle because we are
looking at peak widths rather than standard deviations. The
$sinc^2$ function generated by diffraction actually has an
infinite standard deviation, and alternative measures of
uncertainty and uncertainty relations are therefore required
\cite{uncertainty}.}.

\section{Conditional measurements and the uncertainty principle.}

The uncertainty principle constrains the results of any
measurement of non-commuting observables, and can be represented
by the general inequality \cite{inequality}
\begin{equation} \label{general_uncertainty}
\Delta_{\psi} A \, \Delta_{\psi} B \geqslant \frac{1}{2} |
\bra{\psi} [\hat{A},\hat{B}] \ket{\psi} |,
\end{equation}
which relates the standard deviations of observables $\hat{A}$ and
$\hat{B}$. If $\ket{\psi}$ describes a system of several particles
we can construct an inequality for the position $\hat{x}_n$ and
momentum $\hat{p}_n$ of the $n$'th particle, where $[\hat{x}_n,
\hat{p}_n]=i \hbar$. This yields a familiar Heisenburg uncertainty
relation for each particle in the system, regardless of any
entanglement between them,
\begin{equation} \label{multi-heisenburg}
\Delta x_n \Delta p_n \geqslant \frac{\hbar}{2} \qquad \forall \,
n.
\end{equation}
However, many experiments on entangled systems actually probe
conditional behaviour rather than single-particle properties. In
Kim and Shih's experiment the crucial quantities are \emph{`the
position and momentum of photon 2 in the plane of slit B given
that photon 1 is detected at $D_1$'}. Can we still apply the
uncertainty principle to such conditional quantities?

To investigate this, we will consider a more general case:
\emph{`A measurement $M_2$ of one of two non-commuting observables
$\hat{A}$ or $\hat{B}$ given that a measurement $M_1$ of $\hat{O}$
obtains the result $o$'}.

If $M_1$ precedes $M_2$, the situation is simple. First,
measurement $M_1$ acts on $\ket{\psi}$ according to the projection
postulate, giving
\begin{equation}
\ket{\psi '} = \frac{\hat{P}_o \ket{\psi}}{(\bra{\psi} \hat{P}_o
\ket{\psi})^{\frac{1}{2}}},
\end{equation}
where $\hat{P}_o$ is an operator which projects the system onto
the eigenstate(s) associated with result $o$. The state will then
undergo some unitary evolution into $\ket{\psi''}=\hat{U}
\ket{\psi'}$ before measurement $M_2$. As seen above, the
uncertainty principle is applicable to measurements on any quantum
state, and will therefore constrain the results obtained in
measurement $M_2$ in the normal way, with
\begin{equation} \label{cond_uncertainty}
\Delta A \, \Delta B \geqslant \frac{1}{2} | \bra{\psi''}
[\hat{A},\hat{B}] \ket{\psi''} |.
\end{equation}
If $M_1$ occurs after $M_2$ the situation is more complex, as the
state on which $M_1$ acts will depend on which observable
$\hat{A}$ or $\hat{B}$ was measured in $M_2$ and the result which
was obtained. However, if $\ket{\psi}$ represents a bi-partite
system in which both sub-systems evolve independently, and $M_1$
and $M_2$ act on different sub-systems, then the results of $M_2$
will be bound by the uncertainty relation regardless of the
time-ordering of $M_1$ and $M_2$.

In such cases, the evolution of sub-system 1 is decoupled from
that of sub-system 2 and can be described by the operator
$\hat{P}_{o} \hat{U}_1(\tau)$, where $\hat{U}_1$ is the unitary
time evolution operator for sub-system 1, and $\tau$ is a time
interval containing measurement $M_2$. By tracing $\hat{P}_o$
backwards in time we can rewrite this evolution as
$\hat{U}_1(\tau) \hat{P}_o '$, where $\hat{P}_o '=
\hat{U}_1^{\dag}(\tau) \hat{P}_o \hat{U}_1(\tau)$ represents an
equivalent projection before temporal evolution. A measurement
$M_1$ which obtains the result $o$ \emph{after} measurement $M_2$
can therefore be replaced by a measurement $M_1'$ obtaining $o'$
(with corresponding projector $\hat{P}_o'$) \emph{before}
measurement $M_2$. Because measurements on independent subsystems
can always be shifted through time in this way, their ordering
becomes irrelevant. We can always recast $M_1$ as a measurement
preceding $M_2$, and use the above arguments to apply the
uncertainty principle to the results.

Any conditional measurement in which $M_1$ and $M_2$ act on
separately evolving subsystems will therefore be bound by the
uncertainty principle. In Kim and Shih's experiment these
subsystems are the two photons created during SPDC, which do not
interact with each other after their initial creation and must
evolve independently between measurements when they are space-like
separated. We are specifically interested in a measurement $M_2$
of the position $\hat{A}=\hat{y}$ or momentum $\hat{B}=\hat{p}_y$
of photon 2 in the plane of slit B given that a measurement $M_1$
on photon 1 detects it at $D_1$. According to equation
(\ref{cond_uncertainty}) and the commutation relation $[\hat{y},
\hat{p}_y]=i \hbar$, we expect photon 2 to obey the standard
Heisenburg relation $\Delta y \Delta p_y > \hbar/2$ even though
$y$ and $p_y$ refer to conditional quantities, yet the results of
the experiment appear to violate this relation.

\section{An explanation of the results}

The measured value of $\Delta p_y$ for the conditionally localised
photon is approximately one third that for diffraction at a
physical slit. To obey the uncertainty relation derived above, we
would expect $\Delta y$ to be 2-3 times larger than the slit width
to compensate for its reduced momentum spread. However, we know
that the photon is confined to the unmagnified image of the slit,
so $\Delta y$ should equal the slit width. How can we resolve this
apparent paradox?

The answer lies in our assumption that the image is perfect. If we
consider the blurring introduced by the finite width of the SPDC
source, we find that the image is actually 2-3 times larger than
the physical slit, precisely as predicted by the uncertainty
relation. Narrowing slit B has a noticeable effect because it
picks out the centre of the blurred image (where a perfect image
would lie), and thus alters the spatial confinement of the photon.

The primary factor limiting image resolution is the width of the
region in which photon pairs are created, given by the diameter of
the laser beam pumping the SPDC source ($\approx 3$mm). All photon
trajectories must pass through the laser-pumped region because
this is where photons are created during SPDC, and we should
account for this in our single-photon model. By replacing the SPDC
source with a 3mm aperture only trajectories which pass through
the source region will be carried over to the single-photon
picture, and trajectories which lie outside the source region (and
therefore do not correspond to a valid two-photon trajectory) will
be eliminated. As we will see below, this intuitive step leads to
a blurring of the image which explains the results without
violating the uncertainty relation.

In terms of the SPDC process, we can understand the same effect as
a result of imperfect phase matching. The accuracy of the phase
matching condition $\mathbf{k}_{1}+\mathbf{k}_{2} \simeq 0$ is
limited by the uncertainty relations
\begin{eqnarray} \label{phase_match_eq}
\Delta (k_{1y}+k_{2y}) \Delta y_1 \geqslant \frac{1}{2}, \\ \Delta
(k_{1y}+k_{2y}) \Delta y_2 \geqslant \frac{1}{2},
\end{eqnarray}
derived from equation (\ref{general_uncertainty}). As $\Delta y_1$
and $\Delta y_2$ are limited by the source width, the momenta of
the photons in each pair cannot be precisely anti-correlated, and
representing their trajectories as straight lines through the
source is only an approximation. It is by tracing these straight
lines that we obtain a perfect image, so by disturbing them the
uncertainty tends to cause blurring.

To understand the connection between the two approaches we equate
$k_y=-k_{1y}$ and $k_y'=k_{2y}$ with the wavevector of a photon
entering and leaving the source region respectively. Taking
$\Delta y_1 = \Delta y_2 = \Delta y_s$ as a measure of the source
width, equation (\ref{phase_match_eq}) then becomes $\Delta
(k_y'-k_y) \Delta y_s \geqslant \frac{1}{2}$, which indicates a
diffraction-like disturbance at the source. A photon with a given
incident wavevector $k_y$ will acquire a momentum spread $\Delta
k_y'$ on passing through the source which is characteristic of
diffraction at a slit of width $\Delta y_s$ and is governed by the
uncertainty relation $\Delta k_y' \Delta y_s \geqslant
\frac{1}{2}$. By replacing the SPDC source with an appropriate
slit in the single-photon system we are therefore simulating the
effect of imperfect phase matching. The fact that both the
single-photon and phase-matching approaches yield the same
predictions highlights the power and universality of the
uncertainty principle.

We can estimate the width of the blurred image by treating the
SPDC source as a rectangular aperture in the single-photon system.
Using this simple model each point in the image is spread by
convolution into a $sinc^2$ function of width (between first
minima)
\begin{equation} \label{delta_y_eq}
\Delta y = \frac{2 D \lambda}{s}
\end{equation}
where $D$ is the distance from source to image (745mm), $\lambda$
is the photon wavelength (702.2nm) and $s$ is the source width
(3mm). This gives a blurring of $\Delta y=0.35$mm, which is more
than double the expected width of the image (0.16mm) and is of the
right magnitude to account for the reduction in momentum spread
observed in the results for case (ii). The accuracy of this
analysis can be improved by using a slit profile which better
represents the intensity of photon pair production in the SPDC
source, but the underlying result remains the same; The narrower
the source, the larger the blurring in the image and the greater
$\Delta y$.

The corresponding reduction in $\Delta p_y$ can be explained
geometrically. For photons to travel from a source of width $s$
through a point-like image a distance $D$ ($\gg s$) away their
trajectories must be bounded by the triangular region between the
two, such that
\begin{equation}
\frac{\Delta p_y}{p} \simeq \frac{s}{D}.
\end{equation}
When combined with (\ref{delta_y_eq}) and the De Broglie relation
$p=h/\lambda$ this gives the uncertainty relation $\Delta y \Delta
p_y \approx 2h$ which is characteristic of single-slit diffraction
at a rectangular aperture.

During the free evolution considered above, the momentum
distribution of the photon is conserved. However, if slit B is
narrowed to the same width as slit A (0.16mm) then only those
photons passing through the centre of the blurred image will be
detected at $D_2$. This increased spatial confinement gives the
photons a greater momentum spread and results in a broader pattern
at $D_2$, as observed in the results for case (i).

Interestingly, if $D_1$ detects all photons passing through slit
A, then we will \emph{only} obtain a $sinc^2$ diffraction pattern
in case (i) if the image is blurred. Consider the wavefunction of
the two-photon entangled system when photon 1 is in the jaws of
slit A and photon 2 in the image\footnote{These two events will
not actually occur simultaneously, but as the photons evolve
independently we can consider the wavefunction
$\ket{\psi}=(\hat{U}_1 (t_1) \otimes \hat{U}_2 (t_2))
\ket{\psi_0}$ where the two photons are effectively studied at
different times};
\begin{equation}
\ket{\psi} \propto \int_{-s/2}^{s/2} \ket{y}_1 \ket{-y}_2
\,\mathrm{d}y.
\end{equation}
The state of photon 2 is given by the reduced density matrix
$\hat{\rho}_2=\mathrm{Tr}_1(\ket{\psi} \bra{\psi})$, which is
actually an incoherent mixed state of the image points,
\begin{equation}
\hat{\rho}_2 \propto  \int_{-s/2}^{s/2} \ket{y}_{22} \bra{y}
\,\mathrm{d}y.
\end{equation}
Each point in the image will therefore evolve with an infinite
momentum spread and over the tracking range of $D_2$ results will
be almost constant. It is only when the image points are blurred
into coherent functions which spread over the width of slit B that
a $sinc^2$ interference pattern will be obtained in the results.

Some of the coherence may also be restored if $D_1$ only detects a
subset of the photons passing through slit A. This is a form of
quantum erasure, in which information about which part of slit A
the photon passed through is lost.

\section{conclusions}

In their paper, Kim and Shih correctly claim that their experiment
does not violate the uncertainty principle, but we disagree with
their explanation. They insist that the photons propagating
towards $D_2$ \emph{are} conditionally localised to within $\Delta
y=0.16mm$ by the image of slit A, and that their reduced momentum
spread in case (ii) is not a cause for concern because the
uncertainty principle does not apply:

\begin{quotation}
``A quantum must obey the uncertainty principle but the
``conditional behaviour'' of a quantum in an entangled two
particle system is different. The uncertainty principle is not for
conditional behaviour.''
\end{quotation}

In this paper, we have shown that the results of their experiment
can be explained without this assertion. In fact, the uncertainty
principle can be naturally extended to conditional measurements
whenever they involve separately-evolving sub-systems, as is the
case in their experiment.

As we have shown, blurring of the ``image'' not only explains the
observed results without violating the uncertainty principle, but
is actually \emph{a necessary consequence} of the principle due to
position-momentum uncertainty at the SPDC source. To explain the
diffraction patterns observed we actually require blurring to
introduce coherent superpositions across the slit width, and give
a finite momentum spread. The effect is analogous to that which
blurs point-source images (e.g. the image of a star) in any lens
system with a finite aperture.

These conclusions are also supported by the results of the
previous ``ghost imaging'' experiment of Pittman \emph{et al.}, in
which significant blurring is evident in the image.

\end{document}